\documentclass[]{birkmult}
 \newtheorem{thm}{Theorem}[section]
 \newtheorem{cor}[thm]{Corollary}
 \newtheorem{lem}[thm]{Lemma}
 
 \theoremstyle{definition}
 \newtheorem{defn}[thm]{Definition}
 \theoremstyle{remark}

 \numberwithin{equation}{section}

\begin{document}

%
%
%
%
%
%
%
%
%
\title[Hilbert Lattice Equations]
 {Hilbert Lattice Equations\footnote[0]{{\em Annales Henri Poincar\'e}, 
{\bf 10}, 1335-1358 (2010)}}
\author[Megill]{Norman D.~Megill}

\address{%
Boston Information Group\\
19 Locke Ln.\\
Lexington, MA 02420\\
USA}

\email{nm@alum.mit.edu}

\author[Pavi\v ci\'c]{Mladen Pavi\v ci\'c}
\address{Physics Chair\br
Faculty of Civil Engineering\br
University of Zagreb\br
Zagreb, Croatia}
\email{pavicic@grad.hr}


\date{June 25, 2009}

\begin{abstract}
There are five known classes of lattice equations that hold in every
infinite dimensional Hilbert space underlying quantum systems:
generalised orthoarguesian, Mayet's $\mathcal{E}_A$, Godowski,
Mayet-Godowski, and Mayet's E equations. We obtain
a result which opens a possibility that the first two
classes coincide. We devise new algorithms to generate Mayet-Godowski
equations that allow us to prove that the fourth class properly
includes the third. An open problem related to the last class is
answered.  Finally, we show some new results on the
Godowski lattices characterising the third class of equations.
\end{abstract}

\maketitle

\section{Introduction}
\label{sec:intro}

In 1995, Maria Pia Sol{\`e}r \cite{soler,holl95} proved that
an infinite-dimensional Hilbert space can be recovered from
an orthomodular lattice (OML) together with a small number
of additional conditions, with the only ambiguity being
that its field may be real, complex, or quaternionic.
Specifically, any OML that is complete, is atomic, satisfies
a superposition principle, has height at least 4, and has an
infinite set of mutually orthogonal atoms, completely
determines such a Hilbert space.  This provides us with a
dual, purely lattice-theoretical way to work with the Hilbert
spaces of quantum mechanics. In addition to offering the
potential for new insights, the lattice-theoretical approach
may be computationally efficient for certain kinds of quantum
mechanics problems, particularly if, in the future, we are
able to exploit what may be a ``natural'' fit with quantum
computation.

However, the approach cannot be applied straightforwardly because
unlike the equations (identities) defining OML, the additional
conditions needed to recover Hilbert space are first- and
second-order quantified conditions. Quantified conditions can
complicate computational work: trivially, a computer cannot
scan infinite lattices or an infinite number of lattices
to determine if ``there exists'' and/or ``for all'' conditions
are satisfied; more generally, quantified theorem-proving
algorithms may be needed to achieve rigorous results. Thus it
is desirable to find equations that can partially express some
of these quantified conditions, allowing them to be weakened or
possibly even replaced.  The goal is to get as close as possible
to a purely equational description of $\mathcal{C}(\mathcal{H})$
(the lattice of closed subspaces of a Hilbert space
$\mathcal{H}$), in other words to find smaller and smaller
equational varieties that contain it.

Until 1975, the only lattice equations known to hold in
$\mathcal{C}(\mathcal{H})$ were those defining OML itself.
Then Alan Day discovered that a stronger equation, the orthoarguesian
law, also holds.  There have been several advances since then.
In 2000, Megill and Pavi\v ci\'c \cite{mpoa99} discovered an infinite
family of equations that generalised the orthoarguesian law
and called them {\em generalised orthoarguesian laws}. In
2006 Mayet \cite{mayet06} described a family of equations
$\mathcal{E}_A$, obtained with a technique similar to that used to
derive the generalised orthoarguesian laws, that should
further generalise these laws. In this paper we obtain a result
that opens a possibility that the latter class coincide with the former.

While the previous equations are not related to the states lattices
admit, the other equations are. In 1981 Godowski
\cite{godow} discovered an infinite family of equations derived
by considering states on the lattice.  In 1986, Mayet \cite{mayet86}
generalised (strengthened)
Godowski's equations with a new family, but the examples he gave
were shown by Megill and Pavi\v ci\'c to actually be instances of
Godowski's equations.\ \cite{mpoa99}
In 2006 by Megill and Pavi\v ci\'c \cite{pm-ql-l-hql2} showed the
Mayet-Godowski class to be independent from the Godowski class.
However, an algorithm for generating  Mayet-Godowski equations
has to our knowledge been unknown. We provide such an algorithm
in this paper.

In 2006, Mayet \cite{mayet06} discovered several new series of equations
that hold provided the underlying field of $\mathcal{H}$ is real, complex,
or quaternionic, which are also the ones of interest for quantum
mechanics. Mayet found these by considering vector-valued states on
$\mathcal{C}(\mathcal{H})$ and showed that they were independent of any
of the other equations found so far. In this paper we obtain several
new results on these equations.

To achieve our results cited above, we developed several new
algorithms. The main part of this paper describes the two most
important ones, which are incorporated into the computer
programs that found these results.  The first algorithm
(Section~\ref{sec:states}) determines whether a finite OML admits a
``strong set of states'' (defined below) and if not,
an extension to the algorithm (Section~\ref{sec:gen}) generates a
Mayet-Godowski equation that fails in the input OML but holds in every
Hilbert lattice.  The second algorithm (Section~\ref{sec:dynamic})
enables us to prove whether or not this generated equation is
independent from {\em every} equation in the infinite family found by
Godowski.  This second algorithm also enabled us to find Godowski
lattices of much higher order than before and to show that it is
possible to reduce their original size, therefore speeding up
calculations that make use of them; some of these results are presented
at the end of Section~\ref{sec:dynamic}.

The last part of the paper presents two new results that were partly
assisted by our programs.  In Section~\ref{sec:oa-open}, we show that
an example provided by Mayet from his new family of
orthoarguesian-related equations in fact can be derived from the
generalised orthoarguesian laws, leaving open the problem of whether
this new family has members that are strictly stronger than these laws.
In the Section~\ref{sec:open}, we show the solution to an open problem
posed by Mayet \cite{mayet06} concerning his new families of equations
related to strong sets of Hilbert-space-valued states
\cite{mayet06-hql2}.

\section{Definitions for lattice structures}
\label{sec:def1}

We briefly recall the definitions we will need.  For further
information, see Refs.~\cite{beran,mpoa99,pm-ql-l-hql1,pm-ql-l-hql2}.

\begin{defn}\label{def:lattice}{\rm \cite{birk2nd}}
A {\em lattice} is an algebra
${\rm L}=\langle\mathcal{L}_{\rm O},\cap,\cup\rangle$
such that the following conditions are satisfied for any
$a,b,c\in\mathcal{L}_{\rm O}$: $a\cup b=b\cup a,\ a\cap b=b\cap a,\
(a\cup b)\cup c=a\cup(b\cup c),\ (a\cap b)\cap c=a\cap(b\cap c),\
a\cap (a\cup b)=a,\ a\cup (a\cap b)=a$.
\end{defn}

\begin{thm}\label{th:ordering}{\rm \cite{birk2nd}}
The binary relation $\le$ defined on {\rm L} as
$a\le b\quad {\buildrel{\rm def}\over\Longleftrightarrow}
\quad a=a\cap b$ is a partial ordering.
\end{thm}

\begin{defn}{\rm \cite{birk3rd}}
An {\em ortholattice} {\rm (OL)} is an algebra
$\langle\mathcal{L}_{\rm O},',\cap,\cup,0,1\rangle$
such that $\langle\mathcal{L}_{\rm O},\cap,\cup\rangle$ is a lattice
with unary operation $'$ called {\em orthocomplementation}
which satisfies the following
conditions for $a,b\in\mathcal{L}_{\rm O}$ ($a'$ is called
the {\em orthocomplement} of $a$): $a\cup a'=1,\ a\cap a'=0,\
a\le b\ \Rightarrow\ b'\le a',\ a''=a$
\end{defn}

\begin{defn}\label{def:oml-o}{\rm \cite{pav93,p98}}
An {\em orthomodular lattice} {\rm (OML)} is an {\rm OL}
in which the following condition holds:
\ $a\leftrightarrow b=1\ \ \Leftrightarrow\ \ a=b$, \ \
where \ \ $a\leftrightarrow b=1 \ \ \
{\buildrel{\rm def}\over\Longleftrightarrow}$
\ \ $a\to b=1 \ \ \&\  \ \ b\to a=1$, where \ \
$a\to b\ \ {\buildrel\rm def\over =}\ \ a'\cup(a\cap b)$.
\end{defn}

\begin{defn}\label{def:commut}{\rm \cite{zeman}} We say
that $a$ and $b$ {\em commute} in {\rm OML}, and write $aCb$,
when the following equation holds: $a\cap(a'\cup b)\le b$.
\end{defn}

\begin{defn}\label{def:hl}{\rm \protect{\footnote{For additional
definitions of the terms used in this section see
Refs.~\cite{beltr-cass-book,holl95,kalmb86,mpoa99}.}}}
An orthomodular lattice which satisfies the following
con\-di\-tions is a {\em Hilbert lattice}, {\rm HL}.
\begin{enumerate}
\item {\em Completeness:\/}
The meet and join of any subset of
an {\rm HL} exist.
\item {\em Atomicity:\/}
Every non-zero element in an {\rm HL} is greater
than or equal to an atom. (An atom $a$ is a non-zero lattice element
with $0< b\le a$ only if $b=a$.)
\item {\em Superposition principle:\/}
(The atom $c$
is a superposition of the atoms $a$ and $b$ if
$c\ne a$, $c\ne b$, and $c\le a\cup b$.)
\begin{description}
\item[{\rm (a)}] Given two different atoms $a$ and $b$, there is at least
one other atom $c$, $c\ne a$ and $c\ne b$, that is a superposition
of $a$ and $b$.
\item[{\rm (b)}] If the atom $c$ is a superposition of  distinct atoms
$a$ and $b$, then atom $a$ is a superposition of atoms $b$ and $c$.
\end{description}
\item {\em Minimum height:\/} The lattice contains at least
two elements $a,b$ satisfying: $0<a<b<1$.
\end{enumerate}
\end{defn}

Note that atoms correspond to pure states when defined on the lattice.
We recall that the {\it irreducibility\/} and the {\it covering
property\/} follow from the superposition principle.\
\cite[pp.~166,167]{beltr-cass-book} We also recall that any Hilbert
lattice must contain a countably infinite number of atoms.
\cite{ivertsj}

By Birkhoff's HSP theorem \cite[p.~2]{jipsen}, the family {\rm HL}
is not an equational variety, since a finite sublattice is not an
{\rm HL}.  A goal of studying equations that hold in
{\rm HL} is to find the smallest variety that includes
{\rm HL}, so that the fewest number of of non-equational
(quantified) conditions such as the above will be needed to complete the
specification of {\rm HL}.

\begin{defn}\label{def:state} A {\em state} {\rm (}also called
{\em probability measures} or simply {\em probabilities}
{\rm \cite{kalmb83,kalmb86,kalmb98,maczin}}{\rm )}
on a lattice $\mathcal L$
 is a function $m:{\mathcal L}\longrightarrow [0,1]$
such that $m(1)=1$ and $a\perp b\ \Rightarrow\ m(a\cup b)=m(a)+m(b)$,
where $a\perp b$ means $a\le b'$.
\end{defn}

\begin{lem}\label{lem:state}
The following properties hold for any state $m$:
\begin{eqnarray}
&m(a)+m(a')=1\label{eq:state3}\\
&a\le b\ \Rightarrow\ m(a)\le m(b)\label{eq:state4}\\
&0\le m(a)\le 1\label{eq:state5}\\
&m(a_1)=\cdots=m(a_n)=1\
\Leftrightarrow\ m(a_1)+\cdots+m(a_n)=n\label{eq:state6}\\
&m(a_1\cap\cdots\cap a_n)=1\ \Rightarrow\
m(a_1)=\cdots=m(a_n)=1\label{eq:state7}
\end{eqnarray}
\end{lem}

\begin{defn}\label{def:strong} A set $S$ of
states on $\mathcal L$ is called a {\em strong}
\footnote{Some authors use the term {\em rich} instead of {\em strong},
e.g.\ Ref.~\cite[p.\ 21]{ptak-pulm}.} set of states if
\begin{eqnarray}
(\forall a,b\in{\rm L})([(\forall m \in S)(m(a)=1\ \Rightarrow
\ m(b)=1)]\ \Rightarrow\ a\le b)\,.\quad
\label{eq:st-qm}
\end{eqnarray}
\end{defn}

\begin{thm}\label{th:strong}{\rm \cite{mpoa99}} Every Hilbert
lattice admits a
strong set of states.
\end{thm}

\section{Definitions of equational families related to states}
\label{sec:def2}

First we will define the family of equations found by Godowski,
introducing a special notation for them.  These equations
hold in any lattice admitting a strong set of states
and thus, in particular, any Hilbert lattice. \cite{mpoa99}

\begin{defn}\label{def:god-equiv}Let us call the
following expression the {\em Godowski identity}:
\begin{eqnarray}
a_1{\buildrel\gamma\over\equiv}a_n{\buildrel{\rm def}
\over =}(a_1\to a_2)\cap(a_2\to a_3)\cap\cdots
\cap(a_{n-1}\to a_n)\cap(a_n\to a_1),
\ n=3,4,\dots\label{eq:god-equiv}
\end{eqnarray}
We define $a_n{\buildrel\gamma\over\equiv}a_1$ in the same way with
variables $a_i$ and $a_{n-i+1}$ swapped.
\end{defn}

\begin{thm}\label{th:god-eq} Godowski's equations {\em\cite{godow}}
\begin{eqnarray}
a_1{\buildrel\gamma\over\equiv}a_3
&=&a_3{\buildrel\gamma\over\equiv}a_1
\label{eq:godow3o}\\
a_1{\buildrel\gamma\over\equiv}a_4
&=&a_4{\buildrel\gamma\over\equiv}a_1
\label{eq:godow4o}\\
a_1{\buildrel\gamma\over\equiv}a_5
&=&a_5{\buildrel\gamma\over\equiv}a_1
\label{eq:godow5o}\\
&\dots &\nonumber
\end{eqnarray}
hold in all ortholattices, {\em OL}'s, with strong sets of states.
An {\em OL} to which these equations are added is a variety
smaller than {\em OML}.

\end{thm}

We shall call these equations {\rm $n$-Go} {\rm (}{\rm 3-Go},
{\rm 4-Go}, etc.\/{\rm )}.  We also denote by
{\rm $n$GO} {\rm (}{\rm 3GO}, {\rm 4GO}, etc.\/{\rm )} the
{\rm OL} variety determined by {\rm $n$-Go}, and we call
equation {\rm $n$-Go}  the {\rm $n$GO law}.\footnote{The
equation $n$-Go can also be expressed with $2n$ variables:
$a_1\perp b_1\perp a_2\perp b_2\perp\ldots a_n\perp b_n\perp a_1
\Rightarrow (a_1\cup b_1)\cap\cdots(a_n\cup b_n)\le b_1\cup a_2$,
where $n\ge 3$. We remark that if we set $n=2$, this equation
holds in all OMLs, answering a question in
Ref.~\cite[p.~536]{mayet06-hql2}. This can be seen as follows.
The equation that results from setting $n=2$ in the equation
series of Th.~\ref{th:god-eq} has two variables and is easily
shown to hold in all OMLs.  The proof of Th.~3.19 of
Ref.~\cite{mpoa99}, which converts it to the $2n$-variable
form, involves only OML manipulations.}

Next, we define a generalisation of this family, first described by
Mayet. \cite{mayet85}  These equations also hold in all lattices
admitting a strong set of states, and in particular in all HLs.

\begin{defn}\label{def:gge}
A {\em Mayet-Godowski equation} ({\rm MGE}) is an equality
with $n\ge 2$ conjuncts on each side:
\begin{eqnarray}
t_1 \cap \cdots\cap t_n = u_1 \cap \cdots\cap u_n
\end{eqnarray}
where each conjunct $t_i$ (or $u_i$) is a term consisting of
either a variable or a disjunction of two or more distinct
variables:
\begin{eqnarray}
t_i = a_{i,1}\cup \cdots\cup a_{i,p_i}\qquad \mbox{i.e. $p_i$ disjuncts}\\
u_i = b_{i,1}\cup \cdots\cup b_{i,q_i}\qquad \mbox{i.e. $q_i$ disjuncts}
\end{eqnarray}
and where the following conditions are imposed on the set of variables
in the equation:
\begin{enumerate}
\item{All variables in a given term $t_i$ or $u_i$ are
      mutually orthogonal.}
\item{ Each variable occurs the same number of times on each side of
       the equality.}
\end{enumerate}
\end{defn}

We will call a lattice in which all MGEs hold an MGO; i.e., MGO is the
largest class of lattices (equational variety) in which all MGEs hold.

The following three theorems about MGEs and MGOs are proved in
Ref.~\cite{pm-ql-l-hql2}.

\begin{thm}\label{th:mge}
A Mayet-Godowski equation holds in any ortholattice
 $\mathcal L$
ad\-mit\-ting a strong set of states and thus, in particular,
in any Hilbert lattice.
\end{thm}

\begin{thm}\label{th:mgo-in-ngo}
The family of all Mayet-Godowski equations
includes, in particular, the Godowski equations
{\rm [Eqs.~(\ref{eq:godow3o}),
(\ref{eq:godow4o}),\ldots]}; in other words, the class {\rm MGO}
is included in $n${\rm GO} for all $n$.
\end{thm}

\begin{thm}\label{th:mgo-lt-ngo}
The class {\rm MGO} is properly included in all $n${\rm GO}s, i.e.,
not all {\rm MGE} equations can be deduced from the equations
$n$-{\rm Go}.
\end{thm}

\begin{defn}\label{def:stateeqn}
A {\em condensed state equation} is an abbreviated
representation of an MGE
constructed as follows:  all (orthogonality) hypotheses are discarded,
all meet symbols, $\cap$, are changed to $+$, and all join
 symbols, $\cup$, are changed to juxtaposition.
\end{defn}

For example, the 3-Go equation can be expressed as: \cite{pm-ql-l-hql2}
\begin{eqnarray}
\lefteqn{a\perp d\perp b\perp e\perp c\perp f\perp a
    \qquad\Rightarrow} & & \nonumber \\
& & (a\cup d)\cap (b\cup e)\cap(c\cup f)\ =\
(d\cup b)\cap (e\cup c)\cap(f\cup a),
\label{eq:3gob}
\end{eqnarray}
which, in turn, can be expressed by the condensed state equation
\begin{eqnarray}
ad+be+cf&=&db+ec+fa.\label{eq:3goc}
\end{eqnarray}
The one-to-one correspondence between these
two representations of an MGE should be obvious.

\section{Finding states on finite lattices}
\label{sec:states}

It is possible to express the set of constraints imposed by states
as a   linear programming (LP) problem.  Linear programming is used by
industry to minimise cost, labour, etc., and many efficient programs have
been developed to solve these problems, most of them based on the
simplex algorithm.

We will examine a particular example in detail to illustrate how the
problem is expressed.  For this example we will consider a Greechie
diagram with 3-atom blocks, although the principle is easily extended
to any number of blocks.

If $m$ is a state, then each 3-atom block with atoms $a$, $b$, $c$ and
complements $a'$, $b'$, $c'$ imposes the following constraints:
\begin{eqnarray}
  m(a) + m(b) + m(c)&\ =\ &1 \label{eqn:mmm} \\
  m(a') + m(a)&\ =\ &1       \nonumber\\
  m(b') + m(b)&\ =\ &1       \nonumber\\
  m(c') + m(c)&\ =\ &1       \nonumber\\
  m(x) &\ \ge\ & 0,\qquad x=a,b,c,a',b',c' \nonumber
\end{eqnarray}
To obtain Eq.~(\ref{eqn:mmm}), note that in any Boolean block,
$a \perp b \perp c \perp a$, so $m(a) = 1 - m(a') =
1 - m(b\cup c) = 1 - m(b) - m(c)$.

\begin{figure}[hpt]
\begin{center}
\includegraphics[width=0.3\textwidth]{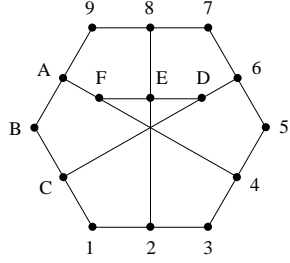}
\end{center}
\caption{Greechie diagram for the Peterson lattice.}
\label{fig:peterson}
\end{figure}

Let us take the specific example of the Peterson lattice, which we
know does not admit a set of strong states.  The Greechie diagram
for this lattice, shown in Fig.~\ref{fig:peterson},
can be expressed with the textual notation

\font\1=cmss8
    {\1123,345,567,789,9AB,BC1,2E8,4FA,6DC,DEF.}

{\parindent=0pt
(see Ref.~\cite{pm-ql-l-hql2}), where each digit or letter represents
an atom, and groups of them represent blocks (edges of the Greechie
diagram).}

Referring to the textual notation,
we designate the atoms by $1,2,\ldots,F$ and their orthocomplements by
$1',2',\ldots,F'$.  We will represent the values of state $m$ on the
atoms by $m(1), m(2), \ldots, m(F)$.  This gives us the following
constraints for the 10 blocks:
\begin{eqnarray}
  m(1) + m(2) + m(3)&\ =\ &1   \nonumber\\
  m(3) + m(4) + m(5)&\ =\ &1   \nonumber\\
  m(5) + m(6) + m(7)&\ =\ &1   \nonumber\\
  m(7) + m(8) + m(9)&\ =\ &1   \nonumber\\
  m(9) + m(A) + m(B)&\ =\ &1   \nonumber\\
  m(B) + m(C) + m(1)&\ =\ &1   \nonumber\\
  m(2) + m(E) + m(8)&\ =\ &1   \nonumber\\
  m(4) + m(F) + m(A)&\ =\ &1   \nonumber\\
  m(6) + m(D) + m(C)&\ =\ &1   \nonumber\\
  m(D) + m(E) + m(F)&\ =\ &1   \nonumber
\end{eqnarray}
In addition, we have $m(a') + m(a) = 1$, $m(a) \ge 0$, and $m(a')\ge 0$
for each atom $a$, adding potentially an additional $15\times 3=45$
constraints.  However, we can omit all but one of these since most
orthocomplemented atoms are not involved this problem, the given
constraints are sufficient to ensure that the state values for atoms are
less than 1, and the particular linear programming algorithm
we  used assumes all variables
are nonnegative.  This speeds up the computation considerably.  The only
one we will need is $m(7) + m(7')=1$ because, as we will see, the
orthocomplemented atom $7'$ will be part of the full problem statement.

We pick two incomparable nodes, 1 and $7'$, which are on opposite sides
of the Peterson lattice.  (The program will try all possible
pairs of incomparable nodes, but for this example we have selected
a priori a pair that
will provide us with the answer.)  Therefore it is the case that $ \sim
1 \le 7'$.  If the Peterson lattice admitted a strong set of states, for
any state $m$ we would have:
\begin{eqnarray}
  (m(1) = 1 \quad\Rightarrow\quad m(7') = 1)
      \quad\Rightarrow\quad 1 \le 7'. \nonumber
\end{eqnarray}

Since the conclusion is false, for some $m$ we must have
\begin{eqnarray}
&&  \sim (m(1) = 1 \quad\Rightarrow\quad  m(7') = 1)      \nonumber\\
&&  {\rm i.e.}\ \ \sim (\sim\ m(1) = 1 \quad\vee\quad m(7') = 1)  \nonumber\\
&&  {\rm i.e.}\ \ \ m(1) = 1 \quad\&\quad \sim\ m(7') = 1         \nonumber
\end{eqnarray}
So this gives us another constraint:
\begin{eqnarray}
  m(1) = 1;       \nonumber
\end{eqnarray}
and for a set of strong states to exist, there must be some $m$ such that
\begin{eqnarray}
  m(7') < 1.     \nonumber
\end{eqnarray}
So, our final linear programming
problem becomes (expressed in the notation of the
publicly available program {\tt lp\_solve}\protect{\footnote{Version 3.2,
available at
{\tt http://groups.yahoo.com/group/lp\_solve/files/}
(as of March 2009).}}):
\begin{verbatim}
  min: m7';
  m1 = 1;
  m7 + m7' = 1;
  m1 + m2 + m3 = 1;
  m3 + m4 + m5 = 1;
  m5 + m6 + m7 = 1;
  m7 + m8 + m9 = 1;
  m9 + mA + mB = 1;
  mB + mC + m1 = 1;
  m2 + mE + m8 = 1;
  m4 + mF + mA = 1;
  m6 + mD + mC = 1;
  mD + mE + mF = 1;
\end{verbatim}
which means ``minimise $m(7')$, subject to constraints $m(1) = 1,
m(7)+m(7')=1, \ldots$.''  The variable to be minimised, $m(7')$, is called
the {\em objective function} (or ``cost function'').  When this problem
is given to {\tt lp\_solve}, it returns an objective function value of
1. This means that regardless of $m$, the other constraints impose a
minimum value of 1 on $m(7')$, contradicting the requirement that $m(7')
< 1$.  Therefore, we have a proof that the Peterson lattice does not
admit a set of strong states.

The program {\tt states.c} that we use reads a list of Greechie diagrams
and, for each one, indicates whether or not it admits a strong set of
states.  The program embeds the {\tt lp\_solve} algorithm, wrapping
around it an interface that translates, internally, each Greechie
diagram into the corresponding linear programming problem.

\section{Generation of MGEs from finite lattices}
\label{sec:gen}

When the linear programming problem in the previous section finds a pair
of incomparable nodes that prove that the lattice admits no strong set
of states, the information in the problem can be used to find an
equation that holds in any OML admitting a strong set of states, and in
particular in HL, but fails in the OML under test.  Typically, an OML to
be tested was chosen because it does not violate any other known HL
equation.  Thus, by showing an HL equation that fails in the OML under
test, we will have found a new equation that holds in HL and is
independent from other known equations.

The set of constraints that lead to the objective function value of 1 in
our linear programming problem turns out to be redundant.  Our algorithm
will try to find a minimal set of hypotheses (constraints) that are
needed.  The equation-finding mode of the {\tt states.c} program
incorporates this algorithm, which will try to weaken the constraints of
the linear programming problem one at a time, as long as the objective
function value remains 1 (as in the problem in the previous section).
The equation will be constructed based on a minimal set of unweakened
constraints that results.

The theoretical basis for the construction is described in the proof of
Theorem 30 of Ref.~\cite{pm-ql-l-hql2}.  Here, we will describe
the algorithm by working through a detailed example.

Continuing from the final linear programming problem of the previous
section, the program will test each constraint corresponding to a
Greechie diagram block, i.e.\ each equation with 3 terms, as follows.
It will change the right-hand side of the constraint equation from $= 1$
to $\le 1$, thus weakening it, then it will run the linear programming
algorithm again.  If the weakened constraint results in an objective
function value $m(7')<1$, it means that the constraint is needed to
prove that the lattice doesn't admit a strong set of states, so we
restore the r.h.s.\ of that constraint equation back to 1. On the other
hand, if the objective function value remains $m(7')=1$ (as in the
original problem), a tight constraint on that block is not needed for
the proof that the lattice doesn't admit a strong set of states, so we
leave the r.h.s.\ of that constraint equation at $\le 1$.

After the program completes this process, the linear programming problem
for this example will look like this:
\begin{verbatim}
min: m7';
m1 = 1;
m7 + m7' = 1;
m1 + m2 + m3 <= 1;
m3 + m4 + m5 = 1;
m5 + m6 + m7 <= 1;
m7 + m8 + m9 <= 1;
m9 + mA + mB = 1;
mB + mC + m1 <= 1;
m2 + mE + m8 = 1;
m4 + mF + mA <= 1;
m6 + mD + mC = 1;
mD + mE + mF <= 1;
\end{verbatim}
Six out of the 10 blocks have been made weaker, and the linear
programming algorithm will show that the objective function has remained
at 1. We now have enough information to construct the MGE, which we will
work with in the abbreviated form of a condensed state equation
(Definition~\ref{def:stateeqn}).
\begin{enumerate}
\item Since $m(1)=1$, the other atoms in the two blocks (3-term
    equations) using it will be $0$.  Thus $m(2)=m(3)=m(B)=m(C)=0$.
\item For each of the four blocks that have $=1$ on the r.h.s., we
    suppress the atoms that are $0$ and juxtapose the remaining 2 atoms in
    each block.  For example, in $m(3)+m(4)+m(5)=1$, we ignore $m(3)=0$, and
    collect the atoms from the remaining two terms to result in $45$ (4
    juxtaposed with 5).  Then we join all four pairs with $+$ to build the
    l.h.s.\ form for the condensed state equation:
    \begin{eqnarray}
      45+9A+E8+6D   \label{eq:lhs}
    \end{eqnarray}
\item For the r.h.s.\ of the equation, we scan the blocks with weakened
    constraints.  From each block, we pick out and juxtapose those
    atoms that also appear on the l.h.s.
    and discard the others.   For example, in $m(5)+m(6)+m(7)\le 1$,
    5 and 6 appear in Eq.~(\ref{eq:lhs}) but 7 doesn't.
    Joining the juxtaposed groups with $+$, we build the r.h.s.:
    \begin{eqnarray}
      56+89+4A+DE    \nonumber
    \end{eqnarray}
    Note that out of the 6 weakened constraints, 2 of them have no
    atoms at all in common with \ Eq.~(\ref{eq:lhs}) and are
    therefore ignored.
\item
    Equating the two sides, we obtain the form of the condensed
     state equation:
    \begin{eqnarray}
      45+9A+E8+6D=56+89+4A+DE  \nonumber
    \end{eqnarray}
\item
    Replacing the atoms with variables, the final condensed state
    equation becomes:
    \begin{eqnarray}
      ab+cd+ef+gh=bg+fc+ad+he  \label{eq:state4go}
    \end{eqnarray}
\item
  Finally, the number of occurrences of each variable on must match on
  each side of the condensed state equation.  In this example, that is
  already the case.  But in general, there may be terms that will have to
  be repeated in order to make the numbers balance.  An example with such
  ``degenerate'' terms is shown as Eq.~(47) of
  Ref.~\cite{pm-ql-l-hql2}.
\end{enumerate}
Eq.~(\ref{eq:state4go}) will be recognised, after converting it to
an MGE, as the 4-Go equation, which as is well-known holds in
all OMLs that admit a strong set of states but fails in the
Peterson lattice Fig.~\ref{fig:peterson}. \cite{mpoa99}

{\bf Remark.} We emphasize that the above algorithm is essentially
heuristic, in that
its purpose is to make use of the existing functions in the {\tt
states.c} program to assist producing new equations
with less manual labor.  In particular, there is no guarantee that it
will produce the strongest equation possible that is deducible from an
OML, nor even that it will be able to find an equation at all.  Indeed,
a few pathological OMLs have been found where the algorithm does not
find terms that can be balanced (in the sense mentioned
in the last item above).

While further refinement of the algorithm may be possible, the point of
it is to provide a practical method to quickly generate new equations
for further study.  These can be independently verified as both holding
in every OML with a strong set of states while at the same time being
stronger than any equation known up to that point.  From a practical
standpoint, at this stage we are mainly interested in studying small
equations with few variables, even if they aren't the strongest possible
that can be generated from an OML, simply because they are more
tractable to work with.  In a similar fashion, many of our results for
$n$-Go and $n$OA equations were obtained by first studying them for
small $n$.  Of course, our goals may change once we gain a better
understanding of the general properties of MGEs and how they can be
classified.

\section{Checking $n$-Go equations on finite lattices}
\label{sec:dynamic}

For the general-purpose checking of whether an equation holds
in a finite lattice, the authors have primarily used a
specialised program, {\tt latticeg.c}, that tests an
equation provided by the user
against a list of Greechie diagrams (OMLs) provided by the user.  This
program has been described in Ref.~\cite{bdm-ndm-mp-1}.  While it has
proved essential to our work, a drawback is that the run time increases
quickly with the number of variables in and size of the input equation,
making it impractical for huge equations.

But there is another limitation in principle, not just in practice, for
the use of the {\tt latticeg.c} program.  In our work with MGEs, we are
particularly interested in those lattices having no strong set of states
but on which all of the successively stronger $n$-Gos pass, for all $n$
less than infinity.  This would prove that any MGE failing in that
lattice is independent from all $n$-Gos and thus represents a new
result.  The {\tt latticeg.c} program can, of course, check only a
finite number of such equations, and when $n$ becomes large the program
is too slow to be practical.  And in any case, it cannot provide a
proof, but only evidence, that a particular lattice does not violate
$n$-Go for any $n$.

Both of these limitations are overcome by a remarkable algorithm based
on dynamic programming, that was suggested by Brendan McKay.  This
algorithm was incorporated into a program, {\tt latticego.c}, that is
run against a set of lattices.  No equation is given to the program;
instead, the program tells the user the first $n$ for which $n$-Go fails
or whether it passes for all $n$.  The program runs very quickly,
depending only on the size of the input lattice, with a run time
proportional to the fourth power of the lattice size (number of nodes)
$m$, rather than increasing exponentially with the equation size (number
of variables) $n$ as with the {\tt latticeg.c} program that checks
against arbitrary equations.

To illustrate the algorithm, we will consider the specific case
of 7-Go.  From this example, the algorithm for the general case
of $n$-Go will be apparent.
We consider $7$-Go written in the following equivalent
 form: \cite{mpoa99}
\begin{eqnarray}
&& (a_1 \to a_2)\cap
 (a_2 \to a_3)\cap
 (a_3 \to a_4)\cap
 (a_4 \to a_5)\cap \nonumber\\
&& \qquad (a_5 \to a_6)\cap
 (a_6 \to a_7)\cap
 (a_7 \to a_1) \ \le\ a_1\to a_7 \label{eq:7go}
\end{eqnarray}
We define intermediate ``operations'' $E_1,\ldots,E_6$ along with
a predicate which provides the answer:
\begin{eqnarray}
&&   E_1(a_1,a_2) = a_1\to a_2  \nonumber\\
&&   E_2(a_1,a_2,a_3) = E_1(a_1,a_2) \cap  (a_2\to a_3)   \nonumber\\
&&   E_3(a_1,a_2,a_3,a_4) = E_2(a_1,a_2,a_3) \cap  (a_3\to a_4)   \nonumber\\
&&   E_4(a_1,a_2,a_3,a_4,a_5) = E_3(a_1,a_2,a_3,a_4) \cap  (a_4\to a_5)
          \nonumber\\
&&   E_5(a_1,a_2,a_3,a_4,a_5,a_6) = E_4(a_1,a_2,a_3,a_4,a_5) \cap
           (a_5\to a_6)  \nonumber\\
&&   E_6(a_1,a_2,a_3,a_4,a_5,a_6,a_7) = E_5(a_1,a_2,a_3,a_4,a_5,a_6) \cap
            (a_6\to a_7)  \nonumber\\
&&   \mbox{answer}(a_1,a_7) = (E_6(a_1,a_2,a_3,a_4,a_5,a_6,a_7)
         \cap  (a_7\to a_1))
           \ \le\ (a_1\to a_7) \nonumber
\end{eqnarray}
Sets of values $V_2,\ldots,V_6$ are computed as follows:
\begin{eqnarray}
&&   V_2(a_1,a_3) = \{E_2(a_1,a_2,a_3) | a_2\}  \nonumber\\
&&   V_3(a_1,a_4) = \{E_3(a_1,a_2,a_3,a_4) | a_2,a_3\}  \nonumber\\
&&   V_4(a_1,a_5) = \{E_4(a_1,a_2,a_3,a_4,a_5) | a_2,a_3,a_4\}  \nonumber\\
&&   V_5(a_1,a_6) = \{E_5(a_1,a_2,a_3,a_4,a_5,a_6) | a_2,a_3,a_4,a_5\}
                 \nonumber\\
&&   V_6(a_1,a_7) = \{E_6(a_1,a_2,a_3,a_4,a_5,a_6,a_7) | a_2,a_3,a_4,a_5,a_6\}
             \nonumber\\
&&   \mbox{For all\ } a_1,a_7:  \mbox{answer}(a_1,a_7) \mbox{\ follows from\ }
              V_6(a_1,a_7), a_7\to a_1, \nonumber\\
&&   \qquad\qquad\mbox{\ and\ } a_1\to a_7   \nonumber
\end{eqnarray}
For example, $V_4(a_1,a_5)$ is the set of values
$E_4(a_1,a_2,a_3,a_4,a_5)$ can have
when $a_2,a_3,a_4$ range over all possibilities.  If
$\mbox{answer}(a_1,a_7)$ is true for all possible $a_1$ and $a_7$,
then 7-Go holds in the lattice, otherwise it fails.

The computation time is estimated as follows, where $m$ is the number
of nodes in the test lattice:
\begin{eqnarray}
&&   \mbox{Each\ } V_2(a_1,a_3)\mbox{\ can be found
       in\ }O(m)\mbox{\ time;\ }O(m^3)\mbox{\ total.} \nonumber\\
&&   \mbox{Each\ } V_3(a_1,a_4)\mbox{\ can be found
         in\ }O(m^2)\mbox{\ time from\ }V_2;
      O(m^4)\mbox{\ total.} \nonumber\\
&&   \mbox{Each\ } V_4(a_1,a_5)\mbox{\ can be found
         in\ }O(m^2)\mbox{\ time from\ }V_3;
      O(m^4)\mbox{\ total.}   \nonumber\\
&&  \qquad\qquad\qquad       \vdots   \nonumber
\end{eqnarray}
So the total time is $O(m^4)$, when the algorithm is applied to a
specific $n$-Go equation.  If it were not for the typical
convergent behavior described below, we would also multiply this time
by $n-2$ i.e. the number of passes $V_2,\ldots,V_{n-1}$.
In fact, only in rare cases do we require a computation of $V_i$
for $i$ greater 10 or so, as we will explain.

The program is written so that it only has to compute additional ``inner
terms'' to process the next $n$-Go equation.  Remarkably, when a lattice
does not violate any $n$-Go, our observation has been that the
addition of new terms almost always converges to a fixed value rather
quickly, meaning that $V_{n}$ for $(n+1)$-Go remains the same as
$V_{n-1}$ for $n$-Go.  This almost always happens for $n<10$, and when
it does, we can terminate the algorithm and say with certainty that no
further increase in $n$ will cause an $n$-Go equation to fail in the
lattice.  (If it doesn't happen, the program will tell us that, but such
a case has so far not been observed.  The program has an arbitrary
cut-off point of $n=100$, after which the algorithm will terminate.  All
of our observed runs have always either converged or failed far below
this point, and in any case the cut-off can be increased with a
parameter setting.)  Convergence provides a proof that the entire class
of Godowski equations (for {\em all} $n<\infty$) will pass in the
lattice.  Such a feat is not possible with ordinary lattice-checking
programs, since an infinite number of equations would have to be tested.

At this point, we do not have a good explanation for this quickly
convergent behavior.  It is simply an empirical observation.

When a lattice does violate some $n$-Go, that result tends to be found
even faster: the algorithm terminates, and the program tells us the
first $n$ at which an $n$-Go equation fails in the lattice.  Since
$n$-Go can be derived from $(n+1)$-Go, failure is also implied for all
greater $n$. The algorithm can also be used for a secondary purpose:  it can
scan a collection of lattices to determine efficiently which of
them satisfy $n$-Go but violate $(n+1)$-Go and find the
smallest ones with this property.

We caution the reader that the $m$ above is lattice nodes, not
Greechie diagram atoms.  For a fixed block size of say 3, which is
the most common one we have used, these numbers are proportional.
However, the number of nodes in a block (Boolean algebra) grows
exponentially with the number of atoms in the block.

\begin{figure}[hpt]
\begin{center}
\includegraphics[width=\textwidth]{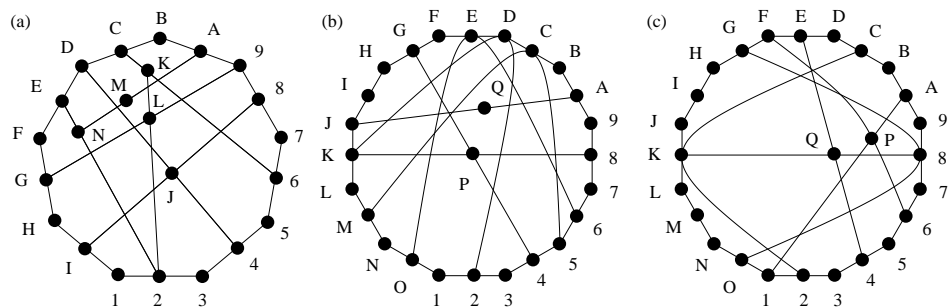}
\end{center}
\caption{(a) 23-16-p7go-f8go-a---one of two smallest lattices
that pass 7-Go and violate 8-Go; (b) 26-18-p8go-f9go-a---one of 23 smallest;
(c) 26-18-p9go-f10go-a---one of 42 smallest.}
\label{fig:8-9-10}
\end{figure}

\begin{figure}[hpt]
\begin{center}
\includegraphics[width=\textwidth]{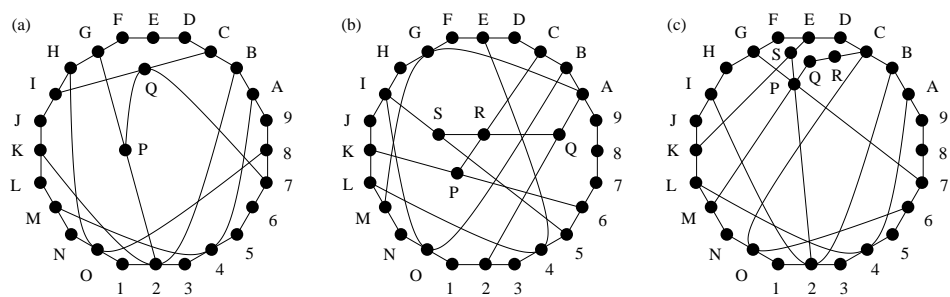}
\end{center}
\caption{(a) 26-18-p9go-f10go-b that also violates $E_3$,
Eq.\ (\ref{eq:E-n}), from Sec.\ \ref{sec:open};
(b) 28-20-p10go-f11go-a---one of the over 50 smallest OMLs
that pass 10-Go and violate 11-Go---it also violates
$E_3$; (c) 28-20-p11go-f12go-a---one of the over 50 smallest OMLs
that pass 11-Go and violate 12-Go---it passes $E_3$.}
\label{fig:9-10-11}
\end{figure}

Here we should explain why we use our algorithm to find
$n$ at which an $n$-Go equation fails in the lattice
when it is well-known \cite{godow} that
$n$-Go fails in Godowski's ``wagon wheel'' of order $n$ which can
be easily constructed for each $n$.
The answer is that smaller lattices significantly reduce the
run time needed to check an equation conjectured to be equivalent
to an $n$-Go.  The $(n+1)$st wagon
wheel lattice has 6 atoms more than the $n$th one, and we have shown
in Ref.\ \cite{mpoa99} that our algorithms give the smallest
lattices in which 4-Go to 7-Go fail that are on average only 2.7
atoms apart.
Our most recent computations\footnote{For this purpose, we used the
{\tt latticego} algorithm
together with the lattice-generating
program {\tt gengre} described in Ref.~\cite{bdm-ndm-mp-1} and
extended in \cite{pmmm03a}.}
presented below show that for higher $n$s,
this number can be still smaller and as we see from
Figs.\ \ref{fig:8-9-10} and  \ref{fig:9-10-11} for, e.g., 9-Go to 12-Go
it is on average 1.

In our textual notation, the OMLs from Fig.\ \ref{fig:8-9-10} read:

{\parindent=0pt
(a) {\1123,\hfil 345,\hfil 567,\hfil 789,\hfil 9AB,\hfil BCD,\hfil DEF,\hfil FGH,\hfil HI1,\hfil 2NE,\hfil 4JD,\hfil 6KC,\hfil IJ8,\hfil GL9,\hfil NMA,\hfil 2LK.\qquad\qquad,\quad}

(b) {\1123,\hfil 345,\hfil 567,\hfil 789,\hfil 9AB,\hfil BCD,\hfil DEF,\hfil FGH,\hfil HIJ,\hfil JKL,\hfil LMN,\hfil NO1,\hfil KP8,\hfil 4PG,\hfil JQA,\hfil OE6,\hfil KD2,\hfil MC5.}

(c) {\1123,\hfil 345,\hfil 567,\hfil 789,\hfil 9AB,\hfil BCD,\hfil DEF,\hfil FGH,\hfil HIJ,\hfil JKL,\hfil LMN,\hfil NO1,\hfil KQ8,\hfil 4QE,\hfil 1PA,\hfil 6PF,\hfil O8G,\hfil 2KC.}}

{\parindent=0pt
The smallest lattice that satisfies 6-Go and violates
7-Go, 24-16-p6go-f7go (24 atoms, 16 blocks) \cite{mpoa99},
is bigger than the smallest one that satisfies 7-Go and
violates 8-Go, 23-16-p7go-f8go (23 atoms, 16 blocks)
shown in Fig.\ \ref{fig:8-9-10}. Also, 26-18-p8go-f9go-a
and 26-18-p9go-f10go-a are of the same size.}

In the textual notation, the OMLs from Fig.\ \ref{fig:9-10-11} (a), (b), and
(c) read:

{\parindent=0pt
{\1123,\hfil 345,\hfil 567,\hfil 789,\hfil 9AB,\hfil BCD,\hfil DEF,\hfil FGH,\hfil HIJ,\hfil JKL,\hfil LMN,\hfil NO1,\hfil 2PG,\hfil IQC,\hfil 7QP,\hfil HO8,\hfil K2B,\hfil M4A.}

{\1123,345,567,789,9AB,BCD,DEF,FGH,HIJ,JKL,LMN,NO1,MGA,IOB,L4E,KP6,IS5,2QA,PRC,QSR.}

{\1123,345,567,789,9AB,BCD,DEF,FGH,HIJ,JKL,LMN,NO1,CO6,I2B,L4A,KSE,MPQ,QRC,2PS,7PG.
}}

We see that 28-20-p10go-f11go-a and 28-20-p11go-f12go-a are
of the same size and contain only two more atoms than
26-18-p9go-f10go-a,b.

\section{Can generalised orthoaguesian equations be
enlarged?}
\label{sec:oa-open}

There is an infinite series of algebraic OML equations
that are apparently at least partly independent of
the conditions that the superpositions and the states OMLs
admit impose on OMLs. In particular, the series properly
overlaps with those characterising states and superpositions,
that have to hold in any Hilbert lattice characterising
quantum systems. A class of such equations are the
so-called {\em generalised orthoarguesian equations} $n${\rm OA}
discovered by Megill and Pavi\v ci\'c. \cite{mpoa99,pm-ql-l-hql2}
We introduce  them as follows.

\begin{defn}
\label{def:noa}
We define an operation
${\buildrel (n)\over\equiv}$ on $n$ variables
$a_1,\ldots,a_n$ ($n\ge 3$) as follows:
\begin{eqnarray}
a_1{\buildrel (3)\over\equiv}a_2\
&{\buildrel\rm def\over =}&\
((a_1\to  a_3)\cap(a_2\to  a_3))
\cup((a_1'\to  a_3)\cap(a_2'\to  a_3)) \\
a_1{\buildrel (n)\over\equiv}a_2\
&{\buildrel\rm def\over =}&\ (a_1{\buildrel (n-1)\over\equiv}a_2)\cup
((a_1{\buildrel (n-1)\over\equiv}a_n)\cap
(a_2{\buildrel (n-1)\over\equiv}a_n))\,,\quad n\ge 4\,.\label{noaoper}
\end{eqnarray}
\end{defn}

\begin{thm}\label{th:noa}
The $n${\rm OA} {\em laws}
\begin{eqnarray}
(a_1\to a_3) \cap (a_1{\buildrel (n)\over\equiv}a_2)
\le a_2\to  a_3\,.\label{eq:noa}
\end{eqnarray}
hold in any {\rm HL}.
\end{thm}

In Ref.~\cite[p.\ 530]{mayet06-hql2}, Mayet attempted to enlarge
the generalised orthoarguesian laws by deriving a family of equations,
$\mathcal{E}_A$, which includes the $n$OA laws as a special
case.  The equations
$\mathcal{E}_A$ are shown to hold in all HLs using a
similar method of proof used to obtain the $n$OA
laws.  In particular, subset relations between subspace sums in
a Hilbert space $\mathcal{H}$ are found by considering sums and
differences of their member vectors. Subspace sums are then
converted to closed subspace joins, using either the the
relation $\mathcal{H}_A+\mathcal{H}_B\subseteq
\mathcal{H}_A\cup\mathcal{H}_B$, which holds in general for for
subspaces $\mathcal{H}_A$ and $\mathcal{H}_B$, or the equation
$\mathcal{H}_A+\mathcal{H}_B= \mathcal{H}_A\cup\mathcal{H}_B$,
which holds whenever $\mathcal{H}_A$ and $\mathcal{H}_B$ are orthogonal.

Mayet gives an example from family
$\mathcal{E}_A$ that is at least as strong as the 3OA
law\footnote{Mayet uses the
notation $OA_{n-2}$ for the equation that we call the
$n$OA law.} in the sense that it implies the latter.
Letting $t_1=c\cup ((a\cup d\cup e)\cap (b\cup
c))$, $t_2=d\cup ((a\cup c\cup e)\cap (b\cup d))$, and $t_3=e\cup
((a\cup c\cup d)\cap (b\cup e))$, Mayet obtains the following
equation:\footnote{Although condition (\ref{eq:mayetEA}) has hypotheses,
it is equivalent to a (closed) equation by \cite[p.~168]{mayet86}, Lemma
1. Therefore we are justified calling it an equation, which we will do
for this and similar conditions.} \cite[p.\ 531]{mayet06-hql2}
\begin{eqnarray}
&& a\perp b \quad\&\quad c\perp d \quad\&\quad d\perp e
        \quad\&\quad c\perp e\ \nonumber\\
&& \qquad  \quad\Rightarrow\quad
  (a\cup b)\cap (c\cup d\cup e)\le
  a\cup (b\cap t_3\cap t_2\cap t_1).
    \label{eq:mayetEA}
\end{eqnarray}
By setting $e=0$ in Eq.~(\ref{eq:mayetEA}), we obtain an equation
that obviously implies the 3OA law, which can be seen when the 3OA law is
expressed in the following 4-variable form (\cite{mpoa99}, Theorem 4.9):
\begin{eqnarray}
a \perp b \ \  \&\ \  c \perp d
\quad \Rightarrow \quad ( a \cup b ) \cap ( c \cup d )\le
a \cup ( b \cap ( c \cup ( ( a \cup d )
\cap ( b \cup c ) ) ) ).\quad \label{eq:3oa-4var}
\end{eqnarray}

However, as we prove in the following theorem, it turns out that
the 3OA law Eq.~(\ref{eq:3oa-4var}) also implies
Eq.~(\ref{eq:mayetEA}), meaning that the latter is not
{\em strictly} stronger than the former; in other words, they
are equivalent.

\begin{thm}\label{th:mayetEA}
An {\rm OML} in which Eq.~(\ref{eq:mayetEA}) holds is a {\rm 3OA} and
vice-versa.
\end{thm}
{\em Proof.}
As we just described, Eq.~(\ref{eq:mayetEA}) implies
the implies the 3OA law.  For the converse, assume that we are given the
3OA law and that the hypotheses
of Eq.~(\ref{eq:mayetEA}) hold.  We obtain three substitution
instances of the 3OA law by
by putting $d\cup e$ for $d$; then $d$, $c\cup e$ for $c$, $d$; then
$e$, $c\cup d$ for $c$, $d$ in Eq.~(\ref{eq:3oa-4var}).
The hypotheses of Eq.~(\ref{eq:3oa-4var}) are then satisfied, and we
conclude, respectively,
\begin{eqnarray}
 ( a \cup b ) \cap ( c \cup d\cup e ) &\le&
    a \cup ( b \cap ( c \cup ( ( a \cup d\cup e )
    \cap ( b \cup c ) ) ) ) \ =\ a \cup ( b \cap t_1) \nonumber  \\
 ( a \cup b ) \cap ( d \cup c\cup e ) &\le&
   a \cup ( b \cap ( d \cup ( ( a \cup c\cup e )
   \cap ( b \cup d ) ) ) ) \ =\ a \cup ( b \cap t_2) \nonumber   \\
 ( a \cup b ) \cap ( e \cup c\cup d ) &\le&
   a \cup ( b \cap ( e \cup ( ( a \cup c\cup d )
   \cap ( b \cup e ) ) ) ) \ =\ a \cup ( b \cap t_3) \nonumber
\end{eqnarray}
Conjoining the right-hand sides,
\begin{eqnarray}
  (a\cup b)\cap (c\cup d\cup e)&\le&
        (a \cup ( b \cap t_1))\cap(a \cup ( b \cap t_2))
             \cap(a \cup ( b \cap t_3)).  \nonumber
\end{eqnarray}
Since $a\perp b$, we have that
$a$ commutes with $b\cap t_i$, $i=1,2,3$, so we can
use the Foulis-Holland theorem (F-H; see e.g.\
\cite[p.~25]{kalmb83}) to apply the distributive law
to the right-hand-side:
\begin{eqnarray}
  (a\cup b)\cap (c\cup d\cup e)&\le&
        (a \cup (( b \cap t_1)\cap( b \cap t_2)
             \cap( b \cap t_3))                     \nonumber \\
     &=&
        (a \cup ( b \cap t_1 \cap t_2 \cap t_3),   \nonumber
\end{eqnarray}
which is the conclusion of Eq.~(\ref{eq:mayetEA}) as required.
\hfill$\Box$

In a word, whether the $\mathcal{E}_A$ equations strictly include
the $n$OA
equations or coincide with them remains an open problem.

\section{Mayet's {E}-equations and a solution to a related
open problem}
\label{sec:open}

A third class of equations makes use of ``vector measures'' \cite{jajte}.
Several new families of equations based on them were
found by Mayet \cite{mayet06}.
He called these measures {\em Hilbert-space-valued states}, defined
as follows.

\begin{defn}\label{def:rh-state} A {\em real
Hilbert-space-valued state}---we
call it an $\mathcal{RH}$
{\em state}---on an orthomodular lattice $\mathcal L$
is a function $s:{\mathcal L}\longrightarrow \mathcal{RH}$,
where $\mathcal{RH}$ is a Hilbert space defined over a real
field, such that (a) $||s(1_{\mathcal L})||=1$,
where $s(a)\in {\mathcal RH}$ is a state vector,
$||s(a)||=\sqrt{(s(a),s(a))}$ is the Hilbert space norm, and
$a\in {\mathcal L}$;
(b) $(\forall a,b\in {\mathcal L})\left[\>a\perp b\
\Rightarrow\ s(a\cup b)=s(a)+
s(b)\>\right]$,
where $a\perp b$ means $a\le b'$;
(c) $(\forall a,b\in {\mathcal L})\left[\>a\perp b\
\Rightarrow\ s(a)\perp s(b)\>\right]$,
where $\>s(a)\perp s(b)\>$ means
the inner product $(s(a),s(b))=0$.
\end{defn}

We also define a subclass of {\rm HL} for which
Def.\ \ref{def:rh-state} will later become relevant:

\begin{defn}\label{def:qhl} A {\em
quantum\/$\,$\footnote{Mayet \cite{mayet06}
calls our family {\rm QHL} by the name
{\em classical Hilbert lattices}, but
since the real and complex fields as well as the quaternion
skew field over which the corresponding Hilbert space is
defined are characteristic of its application in quantum
mechanics, one of us (MP) prefers to call these lattices quantum.
Mayet uses the notation HL for the class we call {\rm QHL}.
He also uses
the notation GHL (generalised Hilbert lattices) for the larger class
defined by omitting the field requirement in
Definition~\ref{def:qhl}, which (see Ref.~\cite[\S\S33,34]{maeda})
is equal to the family we call
{\rm HL} in Definition~\ref{def:hl}.  Finally, the notation
$\mathcal{C}(\mathcal{H})$ is often used to specify the lattice of closed
subspaces of a particular Hilbert space $\mathcal{H}$, typically
when its underlying field is complex
(e.g.\ Ref.~\cite[p.~64]{kalmb83}), in which case
$\mathcal{C}(\mathcal{H})\in\mbox{\rm QHL}$.}
Hilbert lattice}, {\rm QHL}, is a Hilbert lattice orthoiso\-morphic
to the set of closed subspaces of the Hilbert space defined
over either a real field, or a complex field, or a quaternion
skew field.
\end{defn}

The conditions of Lemma~\ref{lem:state} hold when
we replace a real state value $m(a)$ with the square
of the norm of the $\mathcal{RH}$ state value $s(a)$.
In addition, there are number of properties that
hold for $\mathcal{RH}$ states---see \cite{mayet06,pm-ql-l-hql2}.
The following definition of a strong set of $\mathcal{RH}$ states
closely follows Definition \ref{def:strong}, with an essential
difference in the range of the states. We also define a
{\em strong set of $\mathcal{CH}$} and a $\mathcal{QH}$ states.

\begin{defn}\label{def:strong-hs}
A set $S$ of $\mathcal{RH}$ states
$s:{\mathcal L}\longrightarrow {\mathcal{RH}}$ is
called a {\em strong set of $\mathcal{RH}$} states if
\begin{eqnarray}
(\forall a,b\in{\rm L})([(\forall s \in S)(||s(a)||=1\ \Rightarrow
\ ||s(b)||=1)]\ \Rightarrow\ a\le b)\,.\quad
\label{eq:st-rhs}
\end{eqnarray}
\end{defn}

\begin{thm}\label{rh-strong-s}{\rm \cite{mayet06},
      \cite[p.~784]{pm-ql-l-hql2}} Any {\rm QHL}
admits a strong set of $\mathcal{RH}$ states.
\end{thm}
Mayet derives three new families of equations, $E_n$, $E^*_n$,
and $E'_n$, which hold in all HLs but do
not (for $n\ge 3$) hold in all OMLs not admitting strong
sets of Hilbert-space-valued states.\
\cite{mayet06-hql2},
For variables $a_1,\ldots,a_n,b_1,\ldots,b_n,r$, let
$(\mathrm{\Omega})$ be the set of conditions $a_i\perp a_j$ ($i\ne j$) and
$a_i\perp b_i$, for $1\le i,j\le n$.  Define $a=a_1\cup \cdots\cup a_n$,
$b=b_1\cup \cdots\cup b_n$, and $q=(a_1\cup b_1)\cap\cdots\cap(a_n\cup b_n)$.
Then Eqs.\ $E_n$, $E^*_n$, and $E'_n$ ($n\ge 2$) are defined as
\begin{eqnarray}
 (\mathrm{\Omega})\ \quad &\Rightarrow&\quad
          a\cap q \le b,
\label{eq:E-n}\\
 ((\mathrm{\Omega})\ \ \&\ \ r\perp a) \quad &\Rightarrow&\quad
          (a\cup r)\cap q \le b\cup r,
\label{eq:E*-n}\\
 ((\mathrm{\Omega})\ \ \&\ \ r\perp a) \quad &\Rightarrow&\quad
      q \cap (q\to r')\cap(a\cup r)\le b,
\label{eq:mayet-eqs-en-b}
\end{eqnarray}
respectively.
\begin{thm} \label{th:mayet-eqs}
Equations $E_n$ and $E^*_n$ fail in $\mathcal{L}_n$
given in Fig.\ 1 of \cite{mayet06}, $n\ge 3$.
Equation $E'_n$ fails in $\mathcal{L}'_n$
given in Fig.\ 5 of
\cite{mayet06}, $n\ge 3$.
All of these equations
hold in any {\rm OML} with a strong set of
$\mathcal{RH}$ states.
\end{thm}

{\em Proof.}
See Ref.~\cite{mayet06} or
Ref.~\cite[p.~785]{pm-ql-l-hql2} for $E_n$, and
Ref.~\cite{mayet06} for $E^*_n$ and $E'_n$.
(Th.\ 36 in \cite{pm-ql-l-hql2} should be corrected to
read ``$\mathcal{L}_n$'' in place of ``$\mathcal{L}_i$, $i=1,\dots,n$.'')
\hfill$\Box$

The equations of Theorem~\ref{th:mayet-eqs}, which hold in every
{\rm QHL}, do not hold in every {\rm HL}.
They are independent of the modular law and of any $n${\rm OA}
law, $n${\rm GO} law, {\rm MGE}, or combination of them added to the
axioms for {\rm OML}.\  \cite{mayet06},\cite[p.~786]{pm-ql-l-hql2}

As an example, that $E_3$ and $E^*_3$ are independent of the $n$OA
laws for $n=3,4,5$ is shown by the fact that OML L42 (Fig.~7$\>$(b)
from \cite{mpoa99}) satisfies the latter equations but violates
$E_3$ and $E^*_3$. Also, our {\tt states} program shows that L42 has
a strong set of states and thus satisfies all $n$-Go and
MGE equations, showing the
independence of $E_3$ and $E^*_3$ from these. L42 is the smallest
lattice with these properties. Equation $E_3$ does not
fail in lattice $\mathcal{L}_3$ given in Fig.\ 5 of
\cite{mayet06}, showing that $E'_3$ is strictly stronger than $E_3$.

Another example is given in Fig.\ \ref{fig:8-9-10} and
\ref{fig:9-10-11}. OMLs 26-18-p9go-f10go-b and
28-20-p10go-f11go-a violate $E_3$, while
23-16-p7go-f8go-a, 26-18-p8go-f9go-a, 26-18-p9go-f10go-a, and
28-20-p11go-f12go-a satisfy it. All of them satisfy $E_4$.
This OMLs are the only known test on $E_3$ and $E_4$
apart from those mentioned above.

Yet another example is the following Mayet's OML (30 atoms, 19 blocks)
{\parindent=-1pt
{\1123,456,789,ABC,DEF,GHI,JKL,MNO,PQR,STU,147S,ADGT,JMPU,3CL,6FO,9IR,2EQ,5HK,8BN.}}\break
(see Fig.~1 of \cite{navara08}), which satisfies both $E_3$
and $E_4$ as well as all $n$-Go but does not admit any state.

The $E_n$, $E^*_n$, and $E'_n$ families of equations provide us with
additional tools with which to study equations holding in all
{\rm QHL}s but not all {\rm HL}s.  Importantly, they provide
us with a property related to the field of the Hilbert space (and in
particular holding in those Hilbert spaces with the classical fields of
real numbers, complex numbers, and quaternions), something not
previously known to be expressible by an equation.

Mayet showed that $E_n$ follows from $E^*_n$ (by
setting $r=0$),
and he asked \cite[p.~544]{mayet06-hql2} whether $E_n$ and $E^*_n$ are
equivalent.  The answer is affirmative.
\begin{thm}\label{th:mayetE*n}
In any {\rm OML}, equation $E_n$ is a consequence of $E^*_n$ and
vice-versa, for $n \ge 2$.
\end{thm}
{\em Proof.}
We have already seen that $E_n$ follows from $E^*_n$.  For the converse,
assume the hypotheses of $E^*_n$ hold i.e. that
\begin{eqnarray}
  & ((\mathrm{\Omega})\ \&\ r\perp a).
\end{eqnarray}
Starting from the left-hand side of the $E^*_n$ conclusion,
\begin{eqnarray}
\lefteqn{
(a\cup r)\cap q}   \nonumber \\
&=& (a_1\cup \cdots\cup a_n \cup r)
     \cap(a_1\cup b_1)\cap\cdots\cap(a_n\cup b_n) \nonumber \\
&\le& (a_1\cup \cdots\cup a_n \cup r)
     \cap(a_1\cup b_1\cup r\cup r)\cap\cdots\cap(a_n\cup b_n\cup r\cup r)
          \nonumber \\
&=& ((a_1\cup \cdots\cup a_n)
     \cap(a_1\cup b_1\cup r)\cap\cdots\cap(a_n\cup b_n\cup r))\cup r.
       \label{eq:eprf1}
\end{eqnarray}
For the last step above, the distributive law is justified by
F-H, since $r$ commutes with each factor.
Next, we substitute $b_1\cup r$ for $b_1$, \ldots, $b_n\cup r$ for $b_n$
into equation $E_n$.  Since $a_i\perp b_i$ and $a_i\perp r$,
we have $a_i\perp (b_i\cup r)$, so the hypotheses of this substitution
instance of $E_n$ are satisfied.
The conclusion gives
\begin{eqnarray}
(a_1\cup \cdots\cup a_n)
     \cap(a_1\cup (b_1\cup r))\cap\cdots\cap(a_n\cup (b_n\cup r))
& \le & (b_1\cup r)\cup\cdots\cup(b_n\cup r) \nonumber
\end{eqnarray}
which, after rearrangements and joining $r$ to the left-hand-side,
\begin{eqnarray}
((a_1\cup \cdots\cup a_n)
     \cap(a_1\cup b_1\cup r)\cap\cdots\cap(a_n\cup b_n\cup r))\cup r
& \le & b\cup r.
       \label{eq:eprf2}
\end{eqnarray}
Chaining Eqs.~(\ref{eq:eprf1}) and (\ref{eq:eprf2}), we conclude
$E^*_n$.
\hfill$\Box$

Mayet also asked whether an OML exists in which
$E^*_2$ fails.  The answer is negative.

\begin{cor}\label{th:mayetE*2}
Equation $E^*_2$,
\begin{eqnarray}
&& a_1\perp b_1\ \&\ a_2\perp b_2\ \&\ r\perp a_1\ \&\
     a_1\perp a_2\ \&\ a_2\perp r \nonumber\\
&& \qquad  \quad\Rightarrow\quad
  (a_1\cup a_2\cup r)\cap(a_1\cup b_1)\cap(a_2\cup b_2)
  \le b_1\cup b_2\cup r, \label{eq:mayetE*2}
\end{eqnarray}
holds in all {\rm OML}s.
\end{cor}
{\em Proof.}
Mayet showed \cite{mayet06} that
$E_2$ holds in all OMLs, and the previous theorem shows that
$E_2$ and $E^*_2$ are equivalent in any OML.
\hfill$\Box$

We can also generalise the previous Corollary, so as to replace
the orthogonality relations with commutes relations
(Def.~\ref{def:commut}) in all but the fourth hypotheses
of Eq.~(\ref{eq:mayetE*2}).

\begin{cor}\label{th:mayetE*2-2}
The following generalization of Equation $E^*_2$,
\begin{eqnarray}
&& a_1C b_1\ \&\ a_2C b_2\ \&\ rC a_1\ \&\
     a_1\perp a_2\ \&\ a_2C r \nonumber\\
&& \qquad  \quad\Rightarrow\quad
  (a_1\cup a_2\cup r)\cap(a_1\cup b_1)\cap(a_2\cup b_2)
  \le b_1\cup b_2\cup r. \label{eq:mayetE*2strong}
\end{eqnarray}
holds in all {\rm OML}s.
\end{cor}

{\em Proof.}
The proof runs as follows:
\begin{eqnarray}
  \lefteqn{(a_1\cup a_2\cup r)\cap(a_1\cup b_1)\cap(a_2\cup b_2)}   \nonumber\\
  &\  =\ & (a_1\cup b_1)\cap((a_2\cup b_2)\cap(a_2\cup (a_1\cup r)))    \nonumber\\
  &\  =\ & (a_1\cup b_1)\cap(a_2\cup (b_2\cap(a_1\cup r)))    \nonumber\\
  &\  \le\ & (a_1\cup b_1)\cap(a_1'\cup (b_2\cap(a_1\cup r)))    \nonumber\\
  &\  =\ & (a_1\cap(a_1'\cup(b_2\cap(a_1\cup r))))
              \cup (b_1\cap(a_1'\cup(b_2\cap(a_1\cup r))))  \nonumber\\
  &\  \le\ & (a_1\cap(a_1'\cup(b_2\cap(a_1\cup r)))) \cup b_1  \nonumber\\
  &\  \le\ & ((a_1\cup r)\cap(a_1'\cup(b_2\cap(a_1\cup r))) \cup b_1  \nonumber\\
  &\  =\ & ((a_1\cup r)\cap a_1')\cup
           ((a_1\cup r)\cap(b_2\cap(a_1\cup r)))\cup b_1 \nonumber\\
  &\  \le\ & ((a_1\cup r)\cap a_1')\cup b_2 \cup b_1 \nonumber\\
  &\  =\ & ((a_1\cap a_1')\cup(r\cap a_1'))\cup b_2 \cup b_1 \nonumber\\
  &\  =\ & 0\cup(r\cap a_1')\cup b_2 \cup b_1 \nonumber\\
  &\  \le\ & b_1\cup b_2\cup r. \nonumber
\end{eqnarray}
In the second, fourth, seventh, and ninth steps we used F-H, applying
the hypotheses of Eq.~(\ref{eq:mayetE*2strong})
as needed to obtain its prerequisite
commuting conditions.  In the third step we used the hypothesis
$a_1\perp a_2$ i.e.\ $a_2\le a_1'$.  All other steps use simple ortholattice
identities.
\hfill$\Box$

\section{Conclusion}
\label{sec:concl}

In the previous sections, we presented several results obtained in the
field of Hilbert space equations, based on the states defined on the
space.  The idea is to use classes of Hilbert lattice equations for an
alternative representation of Hilbert lattices and Hilbert spaces of
arbitrary quan\-tum systems that might eventually enable a direct
introduction of the states of the systems into quantum computers.  In
applications, infinite classes could then be ``truncated'' to provide us
with finite classes of required length.  The obtained classes would in
turn contribute to the theory of Hilbert space subspaces, which so far
is poorly developed. And it is poorly developed because it turns
out that to describe even the simplest physical system is a
very demanding project.

In 1977, Hultgren and Shimony attempted to describe a spin-1 system
by means of Greechie/Hasse diagrams/lattices using
Stern-Gerlach devices.\ \cite{shimony} Some of their diagrams
were incomplete because, as shown Swift and Wright \cite{anti-shimony},
they did not take into account both electric and magnetic
fields. If they have done so, they could have patched the missing
links in their Fig.\ 3 (dashed lines) and with them their lattice
would read {\tt 123,456,789,ABC,58B}. However, even with that
correction, their description cannot work because Greechie/Hasse
diagrams are not subalgebras of a Hilbert lattice. Lattices
necessary for a complete description of a quantum
system\footnote{Greechie diagrams describe orthogonalities
between one-dimensional sublattices well, but, e.g., spans
of nonorthogonal one-dimensional subspaces cannot be described
by them at all---in the Hilbert space the corresponding
subspaces are {\em not} equal to 1, i.e., they do not
span the whole space while in a Greeche diagram they do.
Consequently a proper lattice of a quantum system must be much
larger than a Greechie diagram, which describes only orthogonalities
between its spin components, because the Hilbert lattice
equations require nonorthogonal atoms to pass a lattice.
More specifically, $n$GO equations fail in Hasse/Greechie
diagrams that describe only orthogonalities of Kochen-Specker
setups but hold in their Hilbert space descriptions as well
as in extended Hilbert lattices that take into account all
needed relations between nonorthogonal atoms and their
joins and meets.\ \cite{bdm-ndm-mp-fresl-09}}
turn out to be too large for a brute-force
approach in which we would first generate all possible
lattices up to a very large number of atoms and then scan
them to extract all required properties. Instead, we adopt
a project of finding efficient algorithms that could enable
us to carry out a partial descriptions of  quantum systems in
the lattice equation approach.

The algorithms and associated computer programs that were developed for
this project were essential to its success.  McKay's dynamic programming
algorithm for $n$-Go equations (Section \ref{sec:dynamic}), together
with its quickly convergent behaviour for large $n$, was particularly
fortuitous.  At the time of its discovery, no other way was known that
could show the independence of the MGE equations from all $n$-Go
equations; at best, only empirical evidence pointing towards that answer
could be accumulated.  Indeed, this problem had remained open for nearly
20 years since Mayet's first publication \cite{mayet85} of these
equations.  Recently, Mayet found a direct proof of this independence
that does not need a computer calculation.  \cite{mayet06-hql2}
Nonetheless, the algorithm still provides a useful tool for testing the
simultaneous validity of all $n$-Go equations for individual OMLs that
are not of the form required by Mayet's theorem.

Thus, there is a strong motivation to find other algorithms that, like
McKay's $n$-Go dynamic programming algorithm, can be applied to other
infinite families, in particular the $n$OA (generalised orthoarguesian)
laws, Eq.~(\ref{eq:noa}).  Assuming similar run-time behaviour could be
achieved, these would provide us with extremely powerful tools that
would let us test finite lattices against the family very quickly
(instead of months or years of CPU time) as well as prove independence
results for the entire infinite family at once (if the valuation set
rapidly converges to a final, fixed value with increasing $n$, as it
does for $n$-Go).

The success of the algorithm for $n$-Go depends crucially on the
structure of a particular representation of the $n$-Go equations, where
variables appear only on one side of the equation and are localised to
an adjacent pair of conjuncts in a chain of conjuncts.  Unfortunately,
all currently known forms of the $n$OA laws have their variables
distributed throughout their (very long) equations.  So another
approach, rather than finding a new algorithm, would be to discover a
new form of the $n$OA laws that better separates their variable
occurrences in such a way that the $n$-Go dynamic programming algorithm
might be applicable.  Both of these approaches are being investigated by
the authors.

In Section~\ref{sec:states}, we described the application of
linear programming to find states on a finite lattice.
The authors are unaware of any previous use of linear programming
methods for this purpose, in particular (for
the present study) determining whether the lattice admits a strong set
of states.  There appear to be relatively few programs that deal with
states, and most of the finite lattice examples in the literature
related to states were found by hand.  A Pascal program written by Klaey
\cite{klaey} is able to find certain kinds of states on lattices, but
for the strong set of states problem it is apparently able only to indicate
``yes'' (if a strong set of states was found) or ``unknown''
otherwise.  The linear programming method provides a definite answer
either way, in the predictable amount of time that the simplex algorithm
takes to run.  Finally, the  linear programming problem
itself (with redundant constraints weakened) provides us with the
information we need to construct a new Hilbert lattice equation that
fails in a given lattice not admitting a strong set of states.

The {\tt states.c} algorithm is actually more general than what we have
described for the present work.  It can also determine whether a finite
lattice admits no states, exactly one state, a full set of states, a
full set of dispersion-free ($\{0,1\}$) states, or group-valued states
on the integers $\mathbb{Z}$.  (Ref.~\cite{navara08}
discusses lattices with some of these properties.)

Mayet's recent and important $E$-equation results \cite{mayet06} provide
us with a powerful new method, the use of Hilbert-space-valued states, to
find previously unknown families of equations that hold in Hilbert
lattices.  For further investigation of these equations, it will be
highly desirable to have a program analogous to our {\tt states.c}
(which works only with real-valued states) that will tell us whether or
not a finite lattice admits a strong set of Hilbert-space-valued states.
This problem seems significantly harder than that of finding real-valued
states, and possible algorithms for doing this are being explored by the
authors.

In Section~\ref{sec:dynamic} (see the 2nd half of the section)
some new computational results on the Godowski lattices
characterising Godowski equations are presented. In particular
we found out that the atom number increase for the successive
smallest lattices in which Godowski equations of order $n$
fail can be reduced from 6---as originally obtained by
Godowski for any $n$---to 1 for $9\le n\le 12$
(see Figs.\ \ref{fig:8-9-10} and  \ref{fig:9-10-11}) and
most probably for all higher $n$s.

In Section~\ref{sec:oa-open}, Theorem~\ref{th:mayetEA} tells us that
further work is needed to determine whether or not Mayet's
$\mathcal{E}_A$ equations are independent of the $n$OA laws.  This
problem is more difficult than it may first appear.  With current
techniques, all that we can do is either prove that that a particular
$\mathcal{E}_A$ equation can be derived from the $n$OA laws, or show
that it is independent only up to some feasibly large $n$.  Unlike the
case with the $n$-Go laws, we have no known algorithm from showing
independence from {\em all} equations in the infinite family of $n$OA
laws.  This open problem stresses the need to find such an algorithm.

In Section~\ref{sec:open}, Theorem~\ref{th:mayetE*n} shows that two of
Mayet's E-equation series, $E_n$, Eq.~(\ref{eq:E-n}) and $E^*_n$,
Eq.~(\ref{eq:E*-n}), are in fact equivalent (in an OML), answering
an open question posed by Mayet.\ \cite[p.~544]{mayet06-hql2}  A third
series, $E'_n$, Eq.~(\ref{eq:mayet-eqs-en-b}), is strictly stronger
than $E_n$ (as already shown in Ref.~\cite[p.~549]{mayet06-hql2}).

The main results of Section~\ref{sec:oa-open} (Theorem~\ref{th:mayetEA})
and Section~\ref{sec:open} (Theorem~\ref{th:mayetE*n}) are both negative
in the sense that they show that equations conjectured to be independent
from others in fact aren't.  Nonetheless, such equations are still
useful in the sense that they provide us with non-obvious new ways to
express the equations they are equivalent to.  In particular, they may
move us a step closer to forms amenable to dynamic programming
algorithms that would test entire infinite families at once.

The programs {\tt latticego.c} and {\tt states.c} described above can
be downloaded from {\tt http://us.metamath.org/\#ql}.

\subsection*{Acknowledgment}
Supported by the {\em Ministry of Science, Education, and
Sport of Croatia} through {\em Distributed Processing and
Scientific Data Visualization} program and {\em Quantum
Computation: Parallelism and Visualization} project
(082-0982562-3160). Computational support was provided by
the cluster {\em Isabella} of the {\em University Computing 
Centre} of the {\em University of Zagreb} and by the {\em 
Croatian National Grid Infrastructure}.


\begin{thebibliography}{10}

\bibitem{soler}
Maria~Pia Sol{\`e}r.
\newblock Characterization of {H}ilbert spaces by orthomodular spaces.
\newblock {\em {\it Comm. Alg.}}, {\bf 23}:219--243, 1995.

\bibitem{holl95}
Samuel~S. {Holland, Jr.}
\newblock Orthomodularity in infinite dimensions; a theorem of {M}.
  {S}ol{\`e}r.
\newblock {\em {\it Bull. Am. Math. Soc.}}, {\bf 32}:205--234, 1995.

\bibitem{mpoa99}
Norman~D. Megill and Mladen Pavi{\v c}i{\'c}.
\newblock Equations, states, and lattices of infinite-dimensional {H}ilbert
  space.
\newblock {\em {\it Int. J. Theor. Phys.}}, {\bf 39}:2337--2379, 2000.

\bibitem{mayet06}
Ren{\'e} Mayet.
\newblock Equations holding in {H}ilbert lattices.
\newblock {\em {\it Int. J. Theor. Phys.}}, {\bf 45}:1216--1246, 2006.

\bibitem{godow}
Radoslaw Godowski.
\newblock Varieties of orthomodular lattices with a strongly full set of
  states.
\newblock {\em {\it Demonstratio Math.}}, {\bf 14}:725--733, 1981.

\bibitem{mayet86}
Ren{\'e} Mayet.
\newblock Equational bases for some varieties of orthomodular lattices related
  to states.
\newblock {\em {\it Algebra Univers.}}, {\bf 23}:167--195, 1986.

\bibitem{pm-ql-l-hql2}
Mladen Pavi{\v c}i{\'c} and Norman~D. Megill.
\newblock {\it Quantum Logic and Quantum Computation}.
\newblock In Kurt Engesser, Dov Gabbay, and Daniel Lehmann, editors, {\em
  Handbook of Quantum Logic and Quantum Structures}, volume~{\it Quantum
  Structures}, pages 751--787. Elsevier, Amsterdam, 2007.

\bibitem{mayet06-hql2}
Ren{\'e} Mayet.
\newblock Ortholattice equations and {H}ilbert lattices.
\newblock In Kurt Engesser, Dov Gabbay, and Daniel Lehmann, editors, {\em
  Handbook of Quantum Logic and Quantum Structures}, volume~{\it Quantum
  Structures}, pages 525--554. Elsevier, Amsterdam, 2007.

\bibitem{beran}
Ladislav Beran.
\newblock {\em Orthomodular Lattices; {A}lgebraic Approach}.
\newblock D. Reidel, Dordrecht, 1985.

\bibitem{pm-ql-l-hql1}
Mladen Pavi{\v c}i{\'c} and Norman~D. Megill.
\newblock {{\it Is Quantum Logic a Logic?}}
\newblock In Kurt Engesser, Dov Gabbay, and Daniel Lehmann, editors, {\em
  Handbook of Quantum Logic and Quantum Structures}, volume~{\it Quantum
  Logic}, pages 23--47. Elsevier, Amsterdam, 2009.

\bibitem{birk2nd}
Garrett Birkhoff.
\newblock {\em Lattice Theory}, volume XXV of {\em American Mathematical
  Society Colloqium Publications}.
\newblock American Mathematical Society, New York, 2nd (revised) edition, 1948.

\bibitem{birk3rd}
Garrett Birkhoff.
\newblock {\em Lattice Theory}, volume XXV of {\em American Mathematical
  Society Colloquium Publications}.
\newblock American Mathematical Society, Providence, Rhode Island, 3rd (new)
  edition, 1967.

\bibitem{pav93}
Mladen Pavi{\v c}i{\'c}.
\newblock Nonordered quantum logic and its {YES}--{NO} representation.
\newblock {\em {\it Int. J. Theor. Phys.}}, {\bf 32}:1481--1505, 1993.

\bibitem{p98}
Mladen Pavi{\v c}i{\'c}.
\newblock Identity rule for classical and quantum theories.
\newblock {\em {\it Int. J. Theor. Phys.}}, {\bf 37}:2099--2103, 1998.

\bibitem{zeman}
J.~Jay Zeman.
\newblock Quantum logic with implications.
\newblock {\em {\it Notre Dame J. Formal Logic}}, {\bf 20}:723--728, 1979.

\bibitem{beltr-cass-book}
Enrico~G. Beltrametti and Gianni Cassinelli.
\newblock {\em The Logic of Quantum Mechanics}.
\newblock Addison-Wesley, 1981.

\bibitem{kalmb86}
Gudrun Kalmbach.
\newblock {\em Measures and Hilbert Lattices}.
\newblock World Scientific, Singapore, 1986.

\bibitem{ivertsj}
P.-A. Ivert and T.~Sj{\"o}din.
\newblock On the impossibility of a finite propositional lattice for quantum
  mechanics.
\newblock {\em {\it Helv. Phys. Acta}}, {\bf 51}:635--636, 1978.

\bibitem{jipsen}
Peter Jipsen and Henry Rose.
\newblock {\em Varieties of Lattices}.
\newblock Springer-Verlag, New York, 1992.

\bibitem{kalmb83}
Gudrun Kalmbach.
\newblock {\em Orthomodular Lattices}.
\newblock Academic Press, London, 1983.

\bibitem{kalmb98}
Gudrun Kalmbach.
\newblock {\em Quantum Measures and Spaces}.
\newblock Kluwer, Dordrecht, 1998.

\bibitem{maczin}
Maciej~J. M{\c a}czy{\'n}ski.
\newblock Hilbert space formalism of quantum mechanics without the {H}ilbert
  space axiom.
\newblock {\em {\it Rep. Math. Phys.}}, {\bf 3}:209--219, 1972.

\bibitem{ptak-pulm}
Pavel Pt{\'a}k and Sylvia Pulmannov{\'a}.
\newblock {\em Orthomodular Structures as Quantum Logics}.
\newblock Kluwer, Dordrecht, 1991.

\bibitem{mayet85}
Ren{\'e} Mayet.
\newblock Varieties of orthomodular lattices related to states.
\newblock {\em {\it Algebra Univers.}}, {\bf 20}:368--396, 1985.

\bibitem{bdm-ndm-mp-1}
Brendan~D. Mc{K}ay, Norman~D. Megill, and Mladen Pavi{\v c}i{\'c}.
\newblock Algorithms for {G}reechie diagrams.
\newblock {\em {\it Int. J. Theor. Phys.}}, {\bf 39}:2381--2406, 2000.

\bibitem{pmmm03a}
Mladen Pavi{\v c}i{\'c}, Jean-Pierre Merlet, Brendan~D. Mc{K}ay, and Norman~D.
  Megill.
\newblock {K}ochen--{S}pecker vectors.
\newblock {\em {\it J. Phys. A}}, {\bf 38}:1577--1592, 2005.
\newblock {\it J. Phys. A\/} {\bf 38}, 3709 (2005).

\bibitem{jajte}
R.~Jajte and A.~Paszkiewicz.
\newblock Vector measures on the closed subspaces of a {H}ilbert space.
\newblock {\em {\it Studia Mathematica}}, {\bf 63}:229--251, 1978.

\bibitem{maeda}
Fumitomo Maeda and Sh{\^u}ichir{\^o} Maeda.
\newblock {\em Theory of Symmetric Lattices}.
\newblock Springer-{V}erlag, New York, 1970.

\bibitem{navara08}
Mirko Navara.
\newblock Small quantum structures with small state spaces.
\newblock {\em {\it Int. J. Theor. Phys.}}, {\bf 47}:36­--43, 2008.

\bibitem{shimony}
Bror~O. {Hultgren, III} and Abner Shimony.
\newblock The lattice of verifiable propositions of the spin-1 system.
\newblock {\em {\it J. Math. Phys.}}, {\bf 18}:381--394, 1977.

\bibitem{anti-shimony}
Arthur~R. Swift and Ron Wright.
\newblock Generalized {S}tern-{G}erlach experiments and the observability of
  arbitrary spin operators.
\newblock {\em {\it J. Math. Phys.}}, {\bf 21}:77--82, 1980.

\bibitem{bdm-ndm-mp-fresl-09}
Mladen Pavi{\v c}i{\'c}, Brendan~D. Mc{K}ay, Norman~D. Megill, and Kre{\v
  s}imir Fresl.
\newblock Reverse {K}ochen-{S}pecker setups.
\newblock {\em {\it In preparation}}, 2009.

\bibitem{klaey}
Matthias Kl{\"a}y.
\newblock {\em Stochastic Models on Empirical Systems, Empirical Logic and
  Quantum Logics, and States on Hypergraphs}.
\newblock PhD thesis, University of {B}ern, {F}aculty of {N}atural {S}cience,
  Fischer {D}ruck, {M}{\"u}nsingen, 1985.

\end{thebibliography}
\end{document}